\documentclass{aadebug}

\corr{0309027}{265}

\begin{document}

\runningheads{Tsuyoshi Ohta et al.}{A mathematical framework}

\title{A mathematical framework for automated bug localization}

\author{
Tsuyosh~Ohta\addressnum{1}\comma\extranum{1},
Tadanori~Mizuno\addressnum{1}
}

\address{1}{
Department of Computer Science, Faculty of Information,
Shizuoka University,
Johoku 3--5--1, Hamamatsu city, 432--8011, Japan
}

\extra{1}{E-mail: \{ohta,mizuno\}@cs.inf.shizuoka.ac.jp}

\pdfinfo{
/Title (A mathematical framework for automated bug localization)
/Author (Tsuyoshi Ohta, and Tadanori Mizuno)
}

\newtheorem{definition}{Definition}
\newcounter{enumcont}

\input{epsf.sty}

\begin{abstract}
In this paper, we propose a mathematical framework for automated bug
localization.
This framework can be briefly summarized as follows.
A program execution can be represented as a rooted acyclic directed
graph.
We define an execution snapshot by a cut-set on the graph.
A program state can be regarded as a conjunction of labels on edges
in a cut-set.
Then we argue that a debugging task is a pruning process
of the execution graph by using cut-sets.
A pruning algorithm, i.e., a debugging task, is also presented.
\end{abstract}


\section{Introduction}
Algorithmic debugging or automated bug localization techniques have been
studied more than two decades.
%
Many individual efforts have been published and implemented, but no
comprehensive and standardized frameworks have been proposed so far.
In this paper, we propose a mathematical framework for 
automated bug localization technique.


\section{A framework for bug localization}


\begin{definition}[execution graph]
An {\em execution graph}
$G = \langle v_0, V, E_d \cup E_c \rangle$
is a rooted acyclic directed graph, which
represents an instance of an execution of a program.
Here, $v_0$, $V$, $E_d$, and $E_c$ are a root vertex, a set of
vertices, a set of data edges, and a set of control edges, respectively.
\end{definition}

The {\em root vertex} represents a start point of the program.
A {\em vertex} in $V$ represents some operation during the
execution, such as an assignment, unification, sending message, etc.
A {\em data edge} is labeled by information that is carried along
with it.
Typically, this edge represents a relation between set/use events on
the same variable and is labeled by a (variable name, value) pair.
A {\em control edge} specifies a relation between a
controlling and a controlled vertices.
For example, a vertex which represents a predicate in {\em if} statement
controls other vertices that denote statements in {\em then} and
{\em else} clauses.
A mapping function on $e$ to its label is denoted as $label(e)$.
A control edge is always labeled as ``true.''

According to the programming paradigm, a program dependence
graph\cite{PDG}, a proof tree, and another similar graph
representation can be employed as a basis of execution graph.

\begin{definition}[cut-set]
In a connected graph $G$ (\/like an execution graph)\/, a {\em cut-set} is a
set of edges whose removal from $G$ leaves $G$ disconnected.
We denote $C = \langle G, G_1, G_2 \rangle$ if a cut-set $C$ cuts a
graph $G$ into two mutually disconnected subgraphs $G_1$ and $G_2$ where
$
C = \{ (v_1, v_2) \in E_d \cup E_c \ |\  v_1 \in G_1, v_2 \in G_2 \}.
$
\end{definition}

\begin{definition}[the order of two cut-sets]
The order of two cut-sets $C_a$ and $C_b$ is defined as follows.
\[
C_a \preceq C_b ~~ \stackrel{def}{=} ~~
G_1^a {~is~a~subgraph~of~} G_1^b
~~~~~~ and ~~~~~~
C_a = C_b ~~ \stackrel{def}{=} ~~ G_1^a {~is~identical~to~} G_1^b.
\]
where $C_a = \langle G, G_1^a, G_2^a \rangle$ and
$C_b = \langle G, G_1^b, G_2^b \rangle$.
The relation $\preceq$ defines a partial order on cut-sets.
\end{definition}

For any given graph, many cut-sets exist.
But only a part of them are allowed for debugging purpose
because such cut-sets must have two important properties:
{\em reproducibility} and {\em stoppability} without any
influence to a program execution.
These properties make problems especially for parallel, concurrent,
or distributed programs which may have data races or deadlocks.

\begin{definition}[state]
For a given cut-set $C$\/, we define a {\em state} of an execution
graph on the $C$ as follows.
\[
S_C = \bigwedge_{e \in C} label(e).
\]
\end{definition}



Intuitively speaking, any program execution can be represented as a
data- and control-flow graph even if the program doesn't written in a
procedural language.
A {\em cut-set} is a mathematical view of a snapshot of the execution.
The order of cut-sets, therefore, shows which snapshot precedes
on the execution.
A {\em state} means a program state to be examined at that snapshot.

\begin{definition}[debugging]
{\em Debugging} is a pruning process of an execution graph. It starts when
a programmer becomes aware one of following phenomena.
\begin{description}
\item[local data anomaly:]
For some data edge $e$\/, $label(e)$ doesn't correspond to that of a
programmer's intention.
\item[local control anomaly:]
For some control edge $e$\/, a programmer concludes the edge shouldn't 
exist.
In other words, an operation on a terminal vertex of the edge shouldn't
have been executed.
\item[global anomaly:]
For a property $A$, like an {\em assertion}\/,
a state $S_C$ on a cut-set $C$ violates it.
This kind of anomaly is well-known as a synchronization error.
{\em Deadlock} is a typical case.
\end{description}
\end{definition}

And the pruning process is as follows.
\begin{enumerate}
\item
On finding a local anomaly,  choose a cut-set $C_e$
which includes the edge identified as the anomaly.
Otherwise, set $C_e$ a cut-set that a programmer has found
the global anomaly on it.
\item
$C_c \leftarrow \{$ all out-edges of root vertex $\}$.
Here, it is obvious that $C_c \preceq C_e$.
\setcounter{enumcont}{\theenumi}
\end{enumerate}

All we have to do is to identify one or more vertices that originate the
anomaly.
Such vertices surely exist between $C_c$ and $C_e$.
Starting with the original execution graph, the following
(a kind of binary search) process successively prunes subgraphs
which never contain culprits of the anomaly.
\begin{enumerate}
\setcounter{enumi}{\theenumcont}
\item \label{looptop}
Choose an appropriate $C_t$ such that $C_c \prec C_t \preceq C_e$.
If a such cut-set doesn't exist (typically, only zero or one vertex exist
between $C_c$ and $C_t$), go to step \ref{finish}.
\item \label{loopbottom}
Examine a state $S_{C_t}$ on $C_t$.
If the state contains one or more anomalies, $C_e \leftarrow C_t$.
Otherwise, $C_c \leftarrow C_t$.
Then go to step \ref{looptop}.
%
\item \label{finish}
If $C_c = C_e$\/~(no vertices exist between two cut-sets),
it means that some indispensable operations are missed at
that execution point.
Otherwise (it means exactly one vertex remains between two cut-sets),
there are two types of culprits on $e \in C_e - C_c$.
\begin{enumerate}
\item \label{resultA}
If $e$ has a local anomaly, an initial vertex of $e$ is the culprit.
Maybe an operation at the vertex is in the wrong.
\item \label{resultB}
Otherwise, $C_e$ must have a global anomaly.
We can find all culprits as:
\begin{enumerate}
\item
$M \leftarrow \phi$ 
\item
for each $a \in S_{C_e}$ do
\begin{enumerate}
\item[]
$M \leftarrow M \cup \{ a \}$
if $S_{C_e - \{ a \}}$ doesn't have the global anomaly.
\end{enumerate}
\end{enumerate}
All initial vertices of edges in $M$ are culprits.
That is to say, such vertices indicate missing {\em critical sections}
starting at that execution points.
\end{enumerate}
\end{enumerate}

%

\section{Related Works}

\begin{figure*}
\hfil
\begin{minipage}[t]{.6\textwidth}
\includegraphics[width=8.5cm]{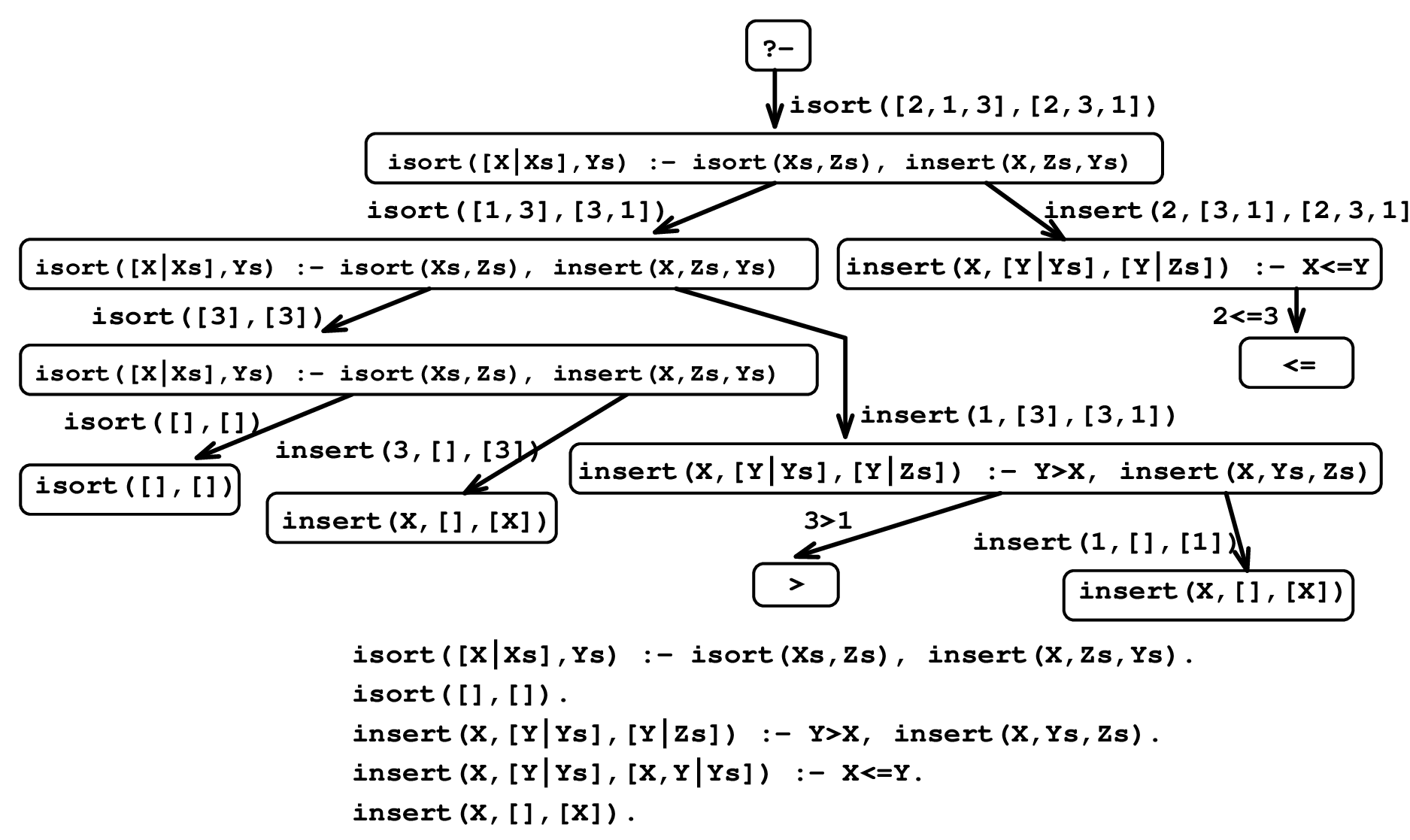}
\caption{Interpretation of Shapiro's method.}
\label{fig:shapiro}
\end{minipage}
\hfil
\begin{minipage}[t]{.35\textwidth}
\includegraphics[width=6cm]{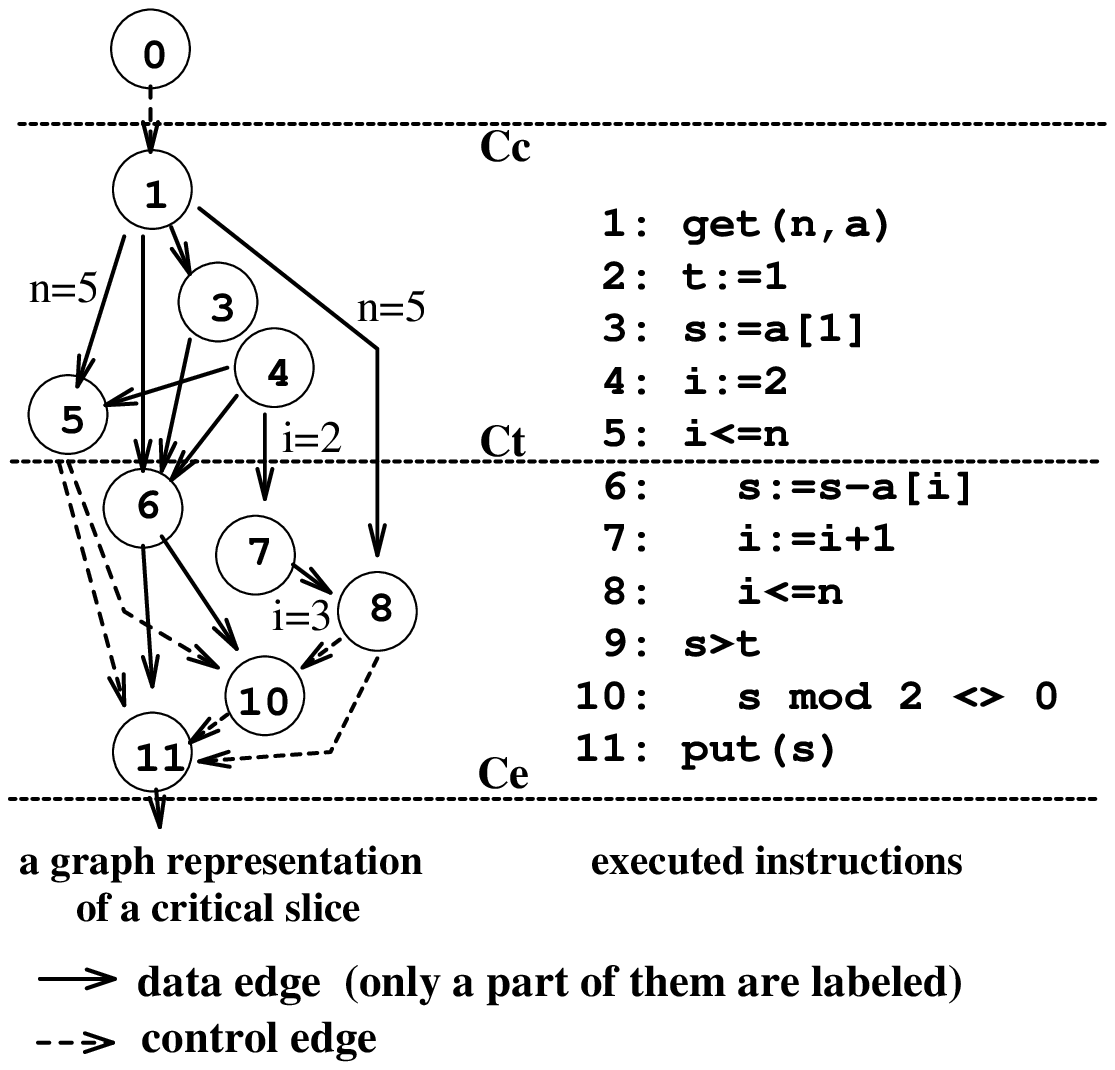}
\caption{Interpretation of Shimomura's method.}
\label{fig:shimomura}
\end{minipage}
\hfil
\end{figure*}

\paragraph*{Shapiro's algorithmic debugging}
was invented for prolog programs\cite{shapiro}.
Fig.~\ref{fig:shapiro} shows our interpretation of his work.
From our viewpoint, it uses a proof tree as an execution graph.
(Attention: This interpretation differs from a normal proof tree.
Our interpretation is based on a line graph\footnote{
A line graph can be get by interchanging vertices and edges of
an original graph.}
of a normal proof tree.)
%
He used only one edge as a cut-set since removal of any edge divides a
tree into two disconnected subtrees. 
A state is also simple because only one label, i.e., one unified clause,
is enough.
In this work, 
step \ref{looptop} of the pruning process is fully automated and
a programmer carries out step \ref{loopbottom} by answering ``yes'' or
``no'' to tell a system the correctness of the label on the edge.
%
GADT\cite{fritzson} and Lichtenstein's system\cite{lichtenstein} can 
be interpreted as the same manner because they are straightforward
extensions of Shapiro's work.

%
%

\paragraph*{\bf FIND}
has developed for sequential procedural languages\cite{shimomuraAA}.
Our interpretation of this work is shown in Fig.~\ref{fig:shimomura}.
It uses an execution graph that represents a {\em critical slice}\/,
which is an extension of dynamic slice\cite{korel}.
A vertex represents a statement execution and an edge represents some relation
between two vertices such as set/use relation of a value of some
variable or control relation of a conditional statement and another statement.
%
FIND uses a traditional breakpoint as a cut-set.
A state was represented as data- and control-flows across the cut-set,
which has ordinary meaning of the word {\em state} we use for procedural
programs.
%
This system carries out step \ref{looptop} automatically and
step \ref{loopbottom} manually.
A programmer examines both data- and control-edges whether they are
correct or not on a cut-set (breakpoint).

\paragraph*{\bf FORMAN}\cite{augustonFORMAN}
also uses a directed graph representing event trace.
It uses two types of edges (relations) between events:
precedence and inclusion.
Compared with our approach, FORMAN has an advantage of modeling power of
hierarchical objects, such as procedure call, with inclusion edges.
But it is too simple for an interactive debugging tool
because precedence edges only models a normal control flow\footnote{
For fairness to FORMAN, it is enough for an off-line event grammar
checker.}.
On the other hand, FORMAN lets an event have attributes to represent
current program status and other things.
So, to represent a program state, FORMAN uses attributes on vertices
while we use a graph structure (a set of labels on edges), i.e.,
a cut-set, due to improving interactive debugging performance.

\paragraph*{\bf Other Approaches:}

From our point of view, constraint or assertion based approaches direct
to automation on step \ref{loopbottom}.
That is to say, their purpose is to check a state without human effort but
using predefined predicates, from which might get a specification of
a program, hopefully.
Knowledge based approach aims at finding better $C_t$
to prune an execution tree as large as possible at one time.
Slicing is a technique to construct an effective
execution tree to find faults.
Here, a word ``effective'' means that edges of the tree lead programmers
to faults as fast as possible without making a detour.

\section{Conclusions}

In this paper, we proposed a mathematical framework for automated bug
localization.
%
%
Based on this framework, we are now implementing an assertion-based
automated bug localization system for distributed programs.
It'll be published near future.

\bibliography{2003}

\end{document}